\documentclass[12pt]{article}
\usepackage{epsfig}
\usepackage{amssymb}
\usepackage{amsmath,amsfonts,amssymb,graphicx}
\newcommand {\be}{\begin{equation}}
\newcommand {\ee}{\end {equation}}
\newcommand{\beq}{\begin{eqnarray}}
\newcommand{\eeq}{\end{eqnarray}}

\begin{document}
\vspace{-2cm}

\title{Time Reversal in Neutrino Oscillations in Matter}
\author{Ernest M. Henley$^{a,b}$, Mikkel B. Johnson$^{c}$, and Leonard S. 
Kisslinger$^{d}$\\ 
 $^a$Department of Physics, University of Washington, Seattle,
WA 98195 \\
 $^b$Institute for Nuclear Theory, University of Washington, 
Seattle, WA 98195\\
 $^c$Los Alamos National Laboratory, Los Alamos, NM 87545\\
 $^d$Department of Physics, Carnegie Mellon University, Pittsburgh, 
PA 15213}
\maketitle
\noindent
PACS Indices:11.30.Er,14.60.Lm,13.15.+g
\vspace{0.25 in}
\begin{abstract}

 We estimate the time reversal violations for neutrino oscillations in
matter for typical experimental energies and baselines. We examine the 
present status of experiments on neutrino oscillations, propose experiments
for TRV, and discuss the future.
\end{abstract}
\vspace {0.25 in}

\section{Introduction}

T and CP violations in neutrino oscillations have long been of interest.
For a review of CP and T violations for neutrino oscillations in vacuum
see, e.g., Ref\cite{bp87}. More than three decades ago the effects of
interactions in matter for neutrino oscillations were estimated\cite{lw78,
ms85}. The effects of matter on T reversal in neutrino oscillations 
have been discussed by a number of authors\cite{1,2}. 

We now briefly review the formalism for T, CP and CPT-violating probability 
differences in neutrino oscillations. Defining the transition probability 
from neutrino a to neutrino b as $\mathcal{P}(\nu_a \rightarrow \nu_b)$,
with flavor a and b = e, $\mu$, or $\tau$.
the  T, CP and CPT probability differences are
\beq
\label{TCPCPT}
      \Delta\mathcal{P}^T_{ab}&=& \mathcal{P}(\nu_a \rightarrow \nu_b)
-\mathcal{P}(\nu_b \rightarrow \nu_a) \nonumber \\
       \Delta\mathcal{P}^{CP}_{ab}&=& \mathcal{P}(\nu_a \rightarrow \nu_b)
-\mathcal{P}(\bar{\nu}_a \rightarrow \bar{\nu}_b) \\
     \Delta\mathcal{P}^{CPT}_{ab}&=& \Delta\mathcal{P}^{CP}_{ba}-
 \Delta\mathcal{P}^T_{ba} \nonumber \; ,
\eeq
where $\bar{\nu}_{a,b}$ is an antineutrino with flavor (a,b). Since 
anti-neutrino oscilations differ from neutrino oscilations 
due to matter effects\cite{jo04}, even though the CPT theorem holds in vacuum,
$\Delta\mathcal{P}^{CPT}_{ab} \neq 0$ for neutrino/antineutrinos in matter,
which we call effective CPT violation.
See Ref\cite{jo04} for references to earlier work on effective CPT violation. 

Since the CPT theorem does not hold for neutrinos traversing matter, the 
relation between T and CP in vacuum\cite{bp87} does not hold, and TR 
violation must be derived separately from CP violation. See Ref\cite{khj11}
for our recent study of CPV. One
objective of the present work is to estimate TRV due to matter effects
for some experiments measuring neutrino oscillation. This is done in
section 2. The main objective of the present work is to propose an experiment
to test T reversal violation. In section 3 the present status of neutrino 
oscillation experiments is reviewed. In section 4 the probability of electron 
to muon conversion is estimated for a typical experiment, which forms the
basis for proposed experiments. 

\section{Background and Estimates of TRV}
  In this section we review the concepts and methods to calculate the neutrino
transition probabilities from which one obtains time reversal probabitities 
associated with neutrino oscillations. We use the notation of 
Ref\cite{2}, most of which is standard. In the unitary transformation, $U$,
defined below, the  basic CP phase is $\delta_{CP}$. As shown in Ref\cite{2}, 
for uniform symmetric matter, which we assume, T reversal violation (TRV) 
vanishes if $\delta_{CP}$=0. See Ref\cite{4} for a discussion of $\delta_{CP}$,
which is not well known. We use a value consistent with those used in other 
studies. 

Neutrinos (and antineutrinos) are produced as $\nu_e, \nu_\mu, \nu_\tau$
together with the named charged leptons. However, neutrinos of definite masses
are $\nu_\alpha$, with $\alpha=1,2,3$. The two forms are connected by a 3 by 3
unitary transformation\cite{3}.
\vspace{.25 cm}

\beq
 U=\left( \begin{array}{lcr} c_{12}c_{13} &s_{12}c_{13} & s_{13}
e^{-i \delta_{CP}} \\
     -s_{12}c_{23}-c_{12}s_{23}s_{13}e^{i\delta_{CP}} & c_{12}c_{23}-s_{12}
s_{23}s_{13}e^{i\delta_{CP}} & s_{23}c_{13} \\
s_{12}s_{23}-c_{12}c_{23}s_{13}e^{i\delta_{CP}} & -c_{12}s_{23}-s_{12}c_{23}
s_{13}e^{i\delta_{CP}} & c_{23}c_{13} \end{array} \right) \nonumber \; ,
\eeq
similar to the CKM matrix for quarks. We have neglected the Majorana phases 
and have used the usual short-hand notation $s_{ij} = sin \theta_{ij}$ and 
$c_{ij} = cos\theta_{ij}$.  We have 
\beq
\nu_a &=& U \nu_\alpha .
\eeq

   As in Eq(1), the TRV probability differences are defined as
\beq
   \Delta\mathcal{P}^T_{ab}&=& \mathcal{P}(\nu_a \rightarrow \nu_b)
-\mathcal{P}(\nu_b \rightarrow \nu_a) \; .
\eeq

It is convenient to use the time evolution matrix, $S(t,t_0)$ to derive
$\Delta\mathcal{P}^T_{ab}$:
\beq
             |\nu(t)> &=& S(t,t_0)|\nu(t_0)> \\
             i\frac{d}{dt}S(t,t_0) &=& H(t) S(t,t_0) \; ,
\eeq
with $H(t)$ the Hamiltonian. In the vacuum
\beq
        S_{ab}(t,t_0)&=& \sum_{j=1}^{3} U_{aj} exp^{i E_j (t-t_0)} U^*_{bj}
\eeq

Since neutrinos travel through matter, we must take into account forward 
charged current neutrino electron scattering in the earth.  The potential 
which describes the interaction is 
\beq
V &=& \sqrt{2} G_F n_e,
\eeq
where $G_F$ is the universal weak interaction Fermi constant, and $n_e$ is 
the density of electrons in matter. Using the matter density $\rho$=3 gm/cc, 
the neutrino-matter potential is $V=1.13 \times 10^{-13}$ eV. Note that
$V \rightarrow -V$ for antineutrinos, the source of effective CPT violation 
in matter.

   The TRV electron-muon probability difference, which is the main topic of 
the present work, is obtained from
\beq
  \Delta\mathcal{P}^T_{e \mu}&=& |S_{21}|^2-|S_{12}|^2
\eeq

With the V included one finds
\beq
     S_{12} &=& c_{23} \beta -is_{23} a A_a \nonumber \\
     S_{21} &=& -(c_{23} \beta +is_{23} a C_a) \nonumber \\
      a &=& s_{13}(\Delta -s_{12} \delta) \; ,
\eeq 
with $ \delta =\delta m_{12}^2/(2 E), \Delta =  \delta m_{13}^2/(2 E)$.
Note that
$\delta \ll \Delta$. We set the CP phase $\delta_{CP}=90^o$. 

  With the approximations $V \le \delta \ll \Delta$,
\beq
     A_a & \simeq & f(t,t_0) I_\alpha*(t,t_0) \nonumber \\
       I_\alpha*(t,t_0)&=& \int_{t_0}^t dt' \alpha^*(t',t)f(t',t) \nonumber \\
     \alpha(t,t_0) &=& cos\omega (t-t_0) -i sin (2\theta) sin \omega (t-t_0)
 \nonumber \\
          f(t,to) &=& e^{-i \Delta (t-t_0)}  \\
 2 \omega &=& \sqrt{\delta^2 + V^2 -2 \delta V cos(2 \theta_{12})} \nonumber \\
         \beta &=& -i sin2\theta sin\omega L \nonumber \\
        C_a &=& A_a \nonumber \;.
\eeq

The angle $\theta$ is defined by
\beq
    cos(2\theta) &=& \frac{\delta cos(2 \theta_{12}) -V}{2 \omega}  \; .
\eeq

From $ \Delta\mathcal{P}^T_{e \mu}= |S_{21}|^2-|S_{12}|^2$ it follows that
\beq
     \Delta\mathcal{P}^T_{e \mu}&=&- 2 s_{13}s_{23}c_{23}(\Delta-s_{12}\delta)
Im[e^{-i \delta_{CP}}\beta^*(A_a-C_a^*)] \; .
\eeq
With the approximations ($V,\delta,\omega \ll \Delta$)\cite{2}, using 
integration by parts it follows (see Ref\cite{2}) that (with $t-t_0\simeq L$)
$I_\alpha*=i(1-cos\omega L e^{-i\Delta L})/\Delta$ + $O(1/\Delta^2$). 
Note that both $\beta$ and $A_a-C_a^*$ are purely imagionary, so 
$Im[\beta^*(A_a-C_a^*)]$ = 0. Therefore there is no TRV for our study
of uniform matter if $\delta_{CP}=0$. We
choose $\delta_{CP} = 90^o$, so $e^{-i \delta_{CP}} = -i$ to simplify
our calculation. Using these approximations it follows that\cite{2}
\beq
 A_a-C_a^*&= & 2iIm[A_a] \simeq i\frac{2}{\Delta}[cos\Delta L- cos\omega L] 
\; ,
\eeq
where $L$ is the baseline length for the neutrino experiment.

From this, and using $\delta \ll \Delta$ it follows that
\beq
\label{deltaT}
    \Delta\mathcal{P}^T_{e \mu}&\simeq&-4 s_{13}s_{23}c_{23}sin\omega L
sin2\theta(cos\Delta L-cos\omega L) \nonumber \\
    &\simeq& 0.374 sin\omega L sin2\theta (cos\omega L-cos\Delta L) 
 \; .
\eeq
where we have used L=t, with the neutrinos having approximately the speed
of light.  We parameters $s_{13}=.187$, and $s_{23}=c_{23}=.707$,
$\theta_{12}=32^o$. With E=1 GeV, $\delta=3.8 \times 10^{-14}$ eV, 
$\Delta=1.2 \times 10^{-12}$ eV, V=$11.3 \times 10^{-14}$ eV, 
$\omega=0.5 \times 10^{-13}$ eV, $sin2\theta =0.342$. Therefore 
$4 s_{13}s_{23}c_{23}sin2\theta$=0.128. We use standard units with 1m=
$5\times10^6$/eV.  Similar relationships for 
$\Delta\mathcal{P}^T_{e \mu}$ have been used by a number of authors studying 
neutrino oscillations. 

With these parameters, using Eq(\ref{deltaT}), we find the magnitude of the 
time reversal violation as a function of L shown in Fig. 1. For E=1 GeV
and the MINOS baseline, as in Fig. 2, $\Delta\mathcal{P}^T_{e \mu}$ is
approximately 3 \%, which could be attained in future experiments if there
were both $\nu_e$ and $\nu_\mu$ beams. 

   Among many experiments studying neutrino oscillation, the MINOS experiments
have covered a wide range of energies\cite{minos}. Since it can only measure 
$\nu_\mu \rightarrow \nu_e$, it  cannot measure TRV, but we evaluate it for
possible future experiments  We now apply Eq(\ref{deltaT}) to evaluate 
$\Delta\mathcal{P}^T_{e \mu}$ for the parameters
relevant to MINOS. The baseline is $L$ = 735 km, and the energy range 
3 to 18 GeV.
\clearpage

 With the other parameters in Eq(\ref{deltaT}) the same as 
those used to obtain Fig. 1, our results are shown in Fig. 2.
\vspace{6cm}

\begin{figure}[ht]
\begin{center}
\epsfig{file=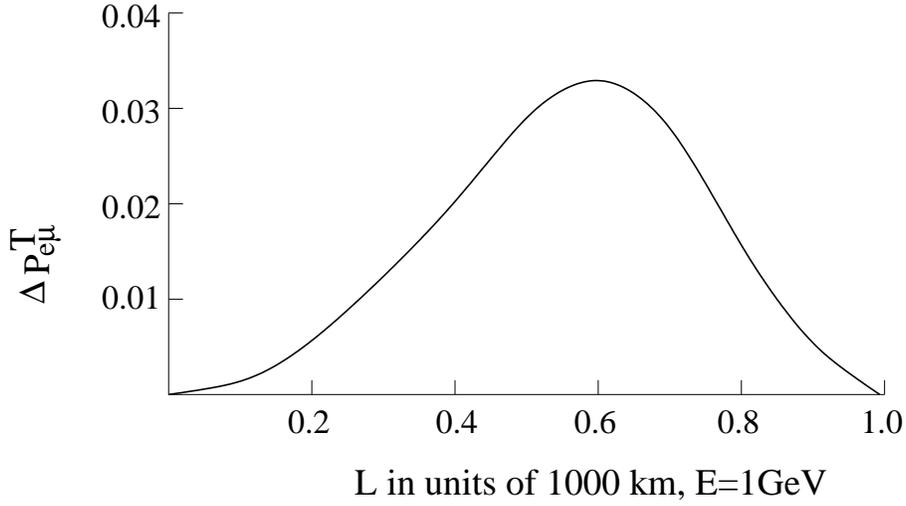,height=2 cm,width=12 cm}
\end{center}
\caption{$\Delta\mathcal{P}^T_{e \mu}$ with E=1 GeV}
\end{figure}
\vspace{6.0cm}

\begin{figure}[ht]
\begin{center}
\epsfig{file=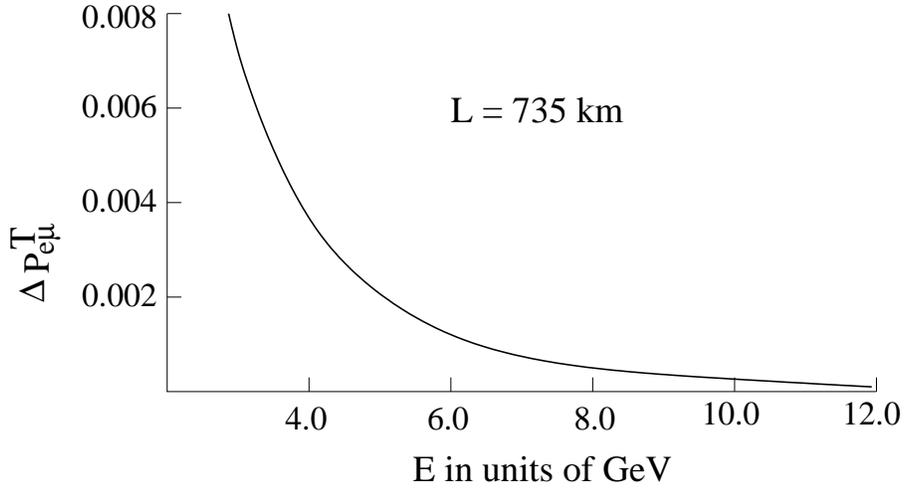,height=2 cm,width=12 cm}
\end{center}
\caption{$\Delta\mathcal{P}^T_{e \mu}$ for L=735 km}
\end{figure}

\newpage

\section {Present Status of Neutrino Oscillation Experiments}

The most likely tests of TRV are those for electron and muon neutrinos or 
antineutrinos. At low energies the conversion of electron antineutrinos has 
been deduced through the disappearance of antineutrinos produced by reactors
\cite {4}. The experiments were carried out in Japan \cite{5}. They are based 
on reactor electron antineutrinos with a mean energy of 3 MeV sent to a 
detector about 180 km away.  The observation of electron neutrino 
disappearance and oscillations was confirmed. 

There are, at present, no definitive muon antineutrino oscillation experiments
to electron antineutrinos. There are some indications for an excess of 
electron neutrinos, over background, in the MINOS experiment, which sends 
muon neutrinos of a few GeV from Fermilab and are detected at the Soudan mine,
735 km away \cite {6}. Similar experiments have been carried out by MiniBooNE 
for both neutrinos and antineutrinos. They report an excess of events in the 
region $ 475 \leq E \leq 1250 MeV$, \cite {7} 
which are consistent with $\bar{\nu}_\mu \rightarrow \bar{\nu}_e$ oscillations
for $ 0.1 \leq \Delta m^2 \leq 1.0 eV^2$. Fits to MiniBooNE and LSND data 
\cite{8} find strong evidence for at least one sterile neutrino\cite{kdcss09,
ppw05,ms07,nw08}.

\section {Proposed Experiment for TRV}

Since there are no sources of high energy electron neutrinos or antineutrinos,
a TRV test via $\bar{\nu}_e \leftrightarrow \bar{\nu}_\mu$ is not possible at 
the present time. In the present work we estimate the probability  
$\mathcal{P} (\nu_e \rightarrow \nu_\mu)$, as a guide for future experiments.
We carry out two sets of calculations, one with a fixed baseline as a function
of energy, and the other with a fixed energy as a function of baseline

\subsection{$\mathcal{P} (\nu_e \rightarrow \nu_\mu)$ for L=735 km}

In this subsection we use the formalism of Ref\cite{4}, with 
 $\mathcal{P} \nu_e \rightarrow \nu_\mu$ given as four terms:
$\mathcal{P} =\mathcal{ P}_0 + \mathcal {P}_{cos \delta_{CP}} 
+\mathcal{P}_{sin \delta_{CP}} + \mathcal{ P}_3$. For the calculation of the
 $\mathcal{P} \nu_e \rightarrow \nu_\mu$  transition probability, with our
choice of $\delta_{CP}$ (see below), the expression is somewhat simpler than 
for the time evolution method used in section 2. Also, we use the MINOS 
baseline, as in section 2; and use parameters 
$\delta m_{31}^2 = 2.4 \times 10^{-3} eV^2$, 
$\delta m_{21}^2 = 7.6 \times 10^{-5} eV^2$; and 
$\theta_{23} = \pi/4$, $\theta_{12} = \pi/5.4$, $\theta_{13}$ 
= 0.188 rad.  Using the notation $A=2 V E$,
with $V$ defined in Eq(8), and $\hat{A}=A/\delta m_{31}^2$, we find (with
$\Delta_L \equiv \Delta L/2$)

\clearpage

\beq
\mathcal{P}_0 &=& sin^2\theta_{23} sin^2(2 \theta_{13}) 
sin^2[(\hat{A}-1) \Delta_L]/(\hat{A} -1)^2, \\
\mathcal{P}_{cos \delta_{CP}} &=& \delta m_{21}^2 cos\delta_{CP} cos 
\theta_{13} sin(2 \theta_{12})sin(2\theta_{13}) sin (2 \theta_{23})\nonumber\\
& & cos \Delta_L sin [\hat{A} \Delta_L] sin 
[(1-\hat{A})\Delta_L]/(A(1-\hat{A})), \\
\mathcal{P}_3 &=& \delta m_{21}^4 cos^2\theta_{23} sin^2(2 \theta_{12}) 
sin^2(\hat{A} \Delta_L)/ A^2,\\
\mathcal{P}_{sin \delta_{CP}} &=& \delta m_{21}^2 sin\delta_{CP} cos 
\theta_{13} sin(2 \theta_{12})sin(2\theta_{13}) sin (2 \theta_{23})\nonumber\\
& & sin \Delta_L sin [\hat{A} \Delta_L] sin 
[(1-\hat{A})\Delta_L]/(A(1-\hat{A})) \; . 
\eeq

As in section 2 we use $\delta_{CP}= 90^o$, so 
sin($\delta_{CP}$)=1.0, cos($\delta_{CP}$)=0.0, and therefore 
$\mathcal{P}_{cos \delta_{CP}}$ = 0. For our estimates of  
$\mathcal{P} \nu_e \rightarrow \nu_\mu$ we use the MINOS baseline, L=735 km, 
as in section 2. 

Because we are not proposing a direct time reversal experiment, e.g., a test
of the equality  $\mathcal{P} (\nu_e \leftrightarrow \nu_\mu)$, but rather 
two experiments which compare these two conversions over the same length L 
at the same energy, we must take all the terms of the probability of
conversion into account.  
Note that the $\mathcal{P}_{sin \delta_{CP}} $ term is positive for 
antineutrinos, but negative for neutrinos. 

From the parameters given above and Eqs(17-20) one finds
\beq
 \mathcal {P}_0 &\simeq& 0.0682, {\rm \;\;} \mathcal {P}_{cos \delta_{CP}}= 0.0,
 \nonumber \\
 \mathcal {P}_3 &\simeq& 0.00073, {\rm \;\;} \mathcal {P}_{sin \delta_{CP}} 
\simeq 0.0186. \nonumber \\
    \mathcal {P}  &\simeq& 0.0875 \nonumber \; .
\eeq
These terms are small because of the small size of $\theta_{13}$

In Fig.3 we give the total $\mathcal{P} \nu_e \rightarrow \nu_\mu$ and
in Fig. 4 the partial probabilities of Eqs(16-29) for L=735 km for
energies appropriate for curent experiments.
\vspace{3cm}

\begin{figure}[htb]
\begin{center}
\epsfig{file=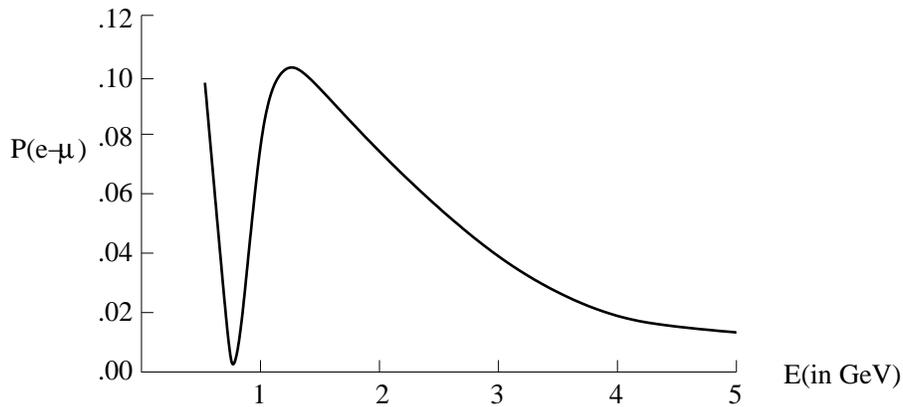,height=2.0cm,width=12cm}
\end{center}
\caption{Overall probability, P, for $\nu_e \rightarrow 
\nu_\mu$ for L=735 km and energies E=0.5 to 5 GeV}
\end{figure}

\clearpage

The partial probabilities are shown in Fig. 4. Since 
$\mathcal{P}_{cos \delta_{CP}}=0$ we do not show that term. 
\vspace{14cm}

\begin{figure}[ht]
\begin{center}
\epsfig{file=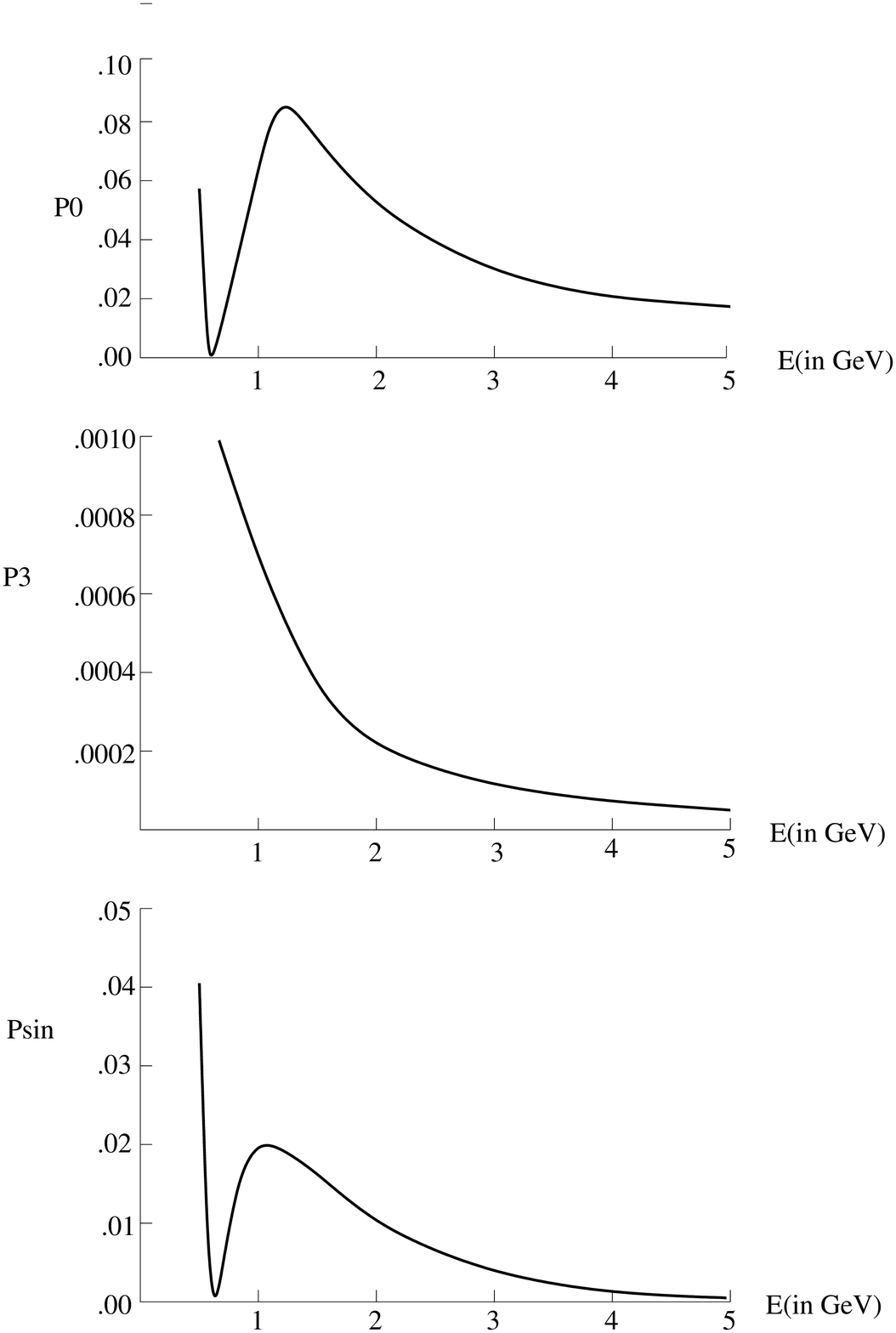,height=2.0cm,width=10cm}
\end{center}
\caption{The partial probabilities for $\nu_e \rightarrow \nu_\mu$ for
L=735 km and energies E=0.5 to 5 GeV}
\end{figure}
 
\clearpage

\subsection{$\mathcal{P} (\nu_e \rightarrow \nu_\mu)$: Accelerator,
 Reactor Experiments}

  In this subsection we find the probability 
$\mathcal{P} (\nu_e \rightarrow \nu_\mu)$ with the baseline and energy
parameters corresponding to the experimental setups MiniBooNE\cite{mini},
MINOS\cite{minos}, JHF-Kamioka\cite{jhf} and CHOOZ\cite{chooz}. MiniBooNE,
MINOS, and JHF-Kamioka have muon neutrino or muon neutrino and muon 
antineutrino beams, while CHOOZ has electron antineutrino beams. The goal 
of this study is to estimate the neutrino conversion probabilities over a 
large range of baselines and energies in order to provide guidance for 
possible TRV experiments in the future.

The results are shown in Fig.5. Note that CHOOZ, with about a 1 km baseline 
and a very low energy has a very large probability. There is a problem, 
however, in identifying the neutrinos or antineutrinos at low energies.

\subsection{ $\mathcal{P} (\nu_e \rightarrow \nu_\mu)$ Matter Effects}

  An important question is how large are matter effects on neutrino conversion
probability. Matter effects are removed by setting $V=0$. In the notation used
in Eqs(17-21), when $V \rightarrow 0$
\beq
     A &\rightarrow& 0 \nonumber \\
     \hat{A} &\rightarrow& 0 \nonumber \\
     \frac{sin[\hat{A} \Delta_L]}{A} &\rightarrow& \frac{\Delta_L}
{\delta m_{31}^2} \;.
\eeq
From this one can show that Eqs(17-20) become

\beq
\mathcal{P}_0 &=& sin^2\theta_{23} sin^2(2 \theta_{13}) sin^2(\Delta_L) \\
\mathcal{P}_{cos \delta_{CP}} &=& (\delta m_{21}^2/\delta m_{31}^2)
cos\delta_{CP} cos \theta_{13} sin(2 \theta_{12})sin(2\theta_{13}) 
sin (2 \theta_{23})\nonumber\\
& & cos(\Delta_L) sin(\Delta_L)\Delta_L , \\
\mathcal{P}_3 &=& (\delta m_{21}^2/\delta m_{31}^2)^2 cos^2\theta_{23} 
sin^2(2 \theta_{12}) \Delta_L^2, \\
\mathcal{P}_{sin \delta_{CP}} &=& (\delta m_{21}^2/\delta m_{31}^2) 
sin\delta_{CP} cos\theta_{13} sin(2\theta_{12})sin(2\theta_{13}) 
sin (2 \theta_{23})\nonumber\\
& &  sin^2(\Delta_L)\Delta_L, 
\eeq

The probabilities $\mathcal{P} (\nu_e \rightarrow \nu_\mu)$ with V=0
are shown as dashed curves in Fig 5. For the short baselines L=500m and
1.03 km, the matter effects are so small we do not show the results for 
V=0.

\clearpage

\begin{figure}[ht]
\begin{center}
\epsfig{file=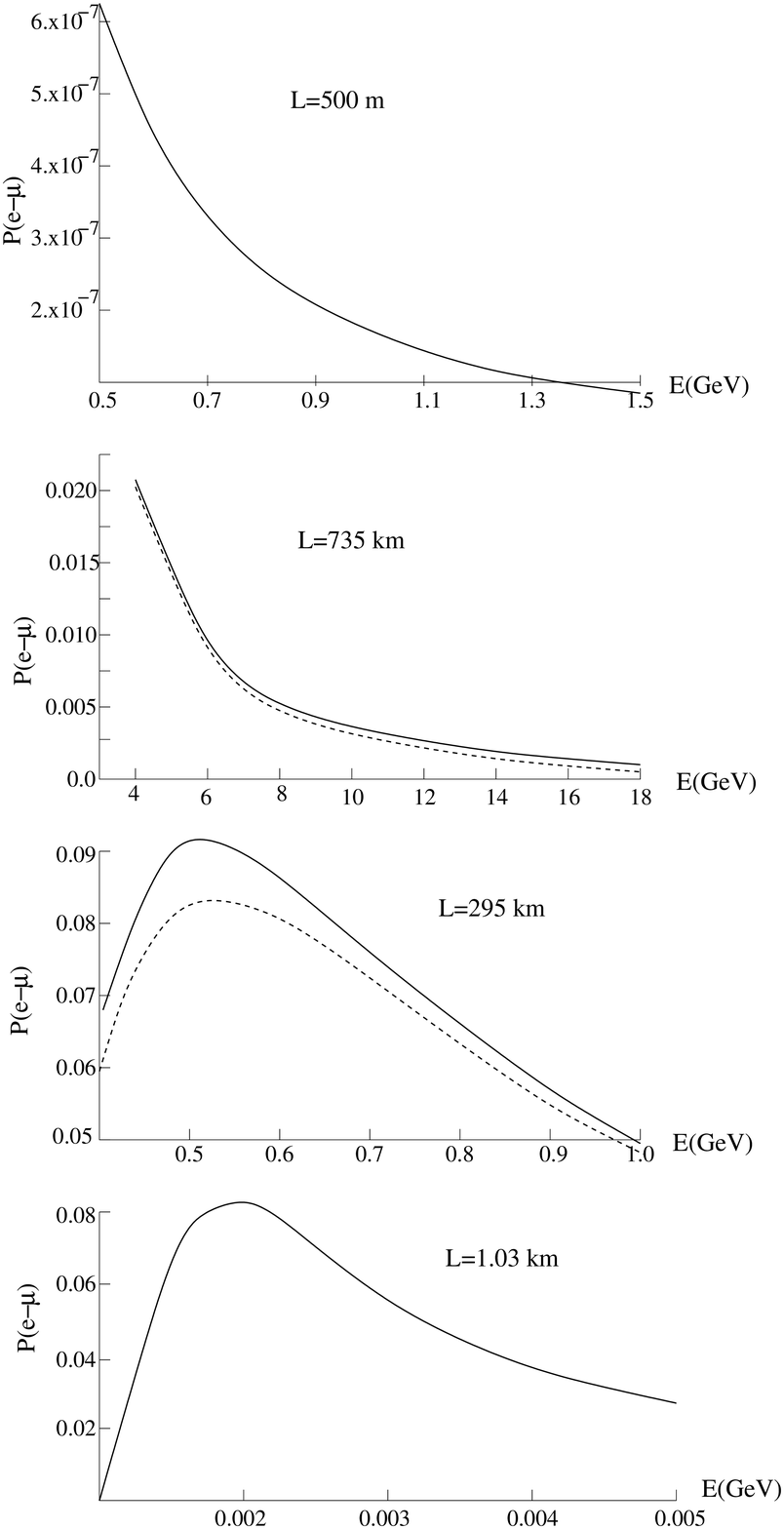,height=16.0cm,width=10cm}
\end{center}
\caption{$\mathcal{P} \nu_e \rightarrow \nu_\mu$ for MiniBooNE(L=500m),
MINOS(L=735km), JHF-Kamioka(L=295km), and CHOOZ (L=1.03km)}
\end{figure}

\clearpage

\section{Conclusions and Proposed Experiments}

In our studies of probability differences, we find a rather
large $\Delta\mathcal{P}^T_{e \mu}$  at 1 GeV with a baseline of 
500 to 700 km (Fig.1), somewhat smaller than the present 
735 km at MINOS, and for the 735 km baseline $\Delta\mathcal{P}^T_{e \mu} 
\simeq .01$ at about 3.0 GeV (Fig.2), within the MINOS range,
if there were both $\nu_e$ and $\nu_\mu$ beams. 

Our studies of the probability of electron to muon neutrino conversion
with parameters corresponding to MiniBooNE, MINOS, JHF-Kamioka, and CHOOZ
find large conversion probablities at low energies, however, identifying
the type of neutrino at low energies is difficult at the present time.
 
 With sufficienly intense antineutrino beams, it might 
be possible to reach an accuracy of $\approx 20\%$. Over time, improvements 
will undoubtedly occur. 

 We conclude with a discussion of specific experiments for future
tests of TRV. At the present time, with only indirect evidence for 
$\bar{\nu}_e \rightarrow \bar{\nu}_\mu$ and no firm evidence for 
$\bar{\nu}_\mu \rightarrow \bar{\nu}_e$, it is difficult to cull data for a 
test of TRI. Our first proposed tests might not be of sufficient accuracy 
to find a TRV, but are, at least steps in the right direction. The tests 
involve two separate experiments, namely 
$\mathcal{P} (\bar{\nu}_e \rightarrow \bar{\nu}_\mu)$ and 
$\mathcal{P}(\bar{\nu}_\mu \rightarrow \bar{\nu}_e)$. We have chosen 
parameters corresponding to those available at accelerators and reactors
at the present time.

  A second proposed test we call the L-2L experiment. 
If $\nu_e \rightarrow \nu_\mu$ with a probability of 10\% at 1 GeV for 
L =735 km (see Fig. 3), then after a further distance of 735 km, if TR 
symmetry holds, 1\% of muon neutrinos will have 
converted back to electron neutrinos.  Assume that 90\% of electron neutrinos 
do not convert at L=735 km and remain $\nu_e$. Then at 1470 km, 81\% of 
electron neutrinos will remain as such, but there will be an added 1\% from the 
conversion to and from muon neutrinos. The difference is small, but may be 
measurable. This argument neglects the conversion of electron neutrinos to 
tau neutrinos. If this is large, it can be used instead of muon neutrinos.
This might be a possible future experiment for MINOS, as well as Kamioka and 
CHOOZ.

\vspace{2 cm}

\Large{{\bf Acknowledgements}}\\
\normalsize
This work was supported in part by the NSF grant PHY-00070888, in part 
by the DOE contracts W-7405-ENG-36 and DE-FG02-97ER41014.


\newpage

\begin{thebibliography}{99}
\bibitem{bp87} S. M. Bilenky and S. T. Petkov, Rev. Mod. Phys. {\bf 59}, 671
(1987)
\bibitem{lw78} L. Wolfenstein, Phys. Rev {\bf D 17}, 2369 (1978) 
\bibitem{ms85} S. P. Mikheev and A. Yu. Smirnov, Sov. J. Nucl. Phys. {\bf 42},
913 (1985)
\bibitem {1} J. Arafune and J. Sato, Phys. Rev {\bf D 55} ,1653 (1997); 
J. Bernab\'{e}u arXiv 9904474; J. Bernab\'{e}u and M.C. Ba\~{n}uls, 
arXiv 0003288 (2000); H. Minakata, H. Nunokawa, and S. Parke, arXiv 0306221.
\bibitem{2} E. Akhmedov, P. Huber, M. Lindner, and T. Ohlsson,  Nucl. Phys. 
{\bf{B608}}, 394 (2001)
\bibitem{jo04} M. Jacobson and T. Ohlsson, Phys. Rev {\bf D 69}, 013003 (2004)
\bibitem{3}See, e.g., J. Arafune and J. Sato, Phys. Rev {\bf D 55}, 1653 (1997)
\bibitem{khj11} L.S. Kisslinger, E.M. Henley, and M.B. Johnson,
arXiv:1105.2741/hep-ph (2011) 
\bibitem{4} M. Freund, Phys. Rev. {\bf{D64}}, 053003 (2001)
\bibitem{minos}MINOS Collaboration, Nucl Phys. B (Proc Supl) {\bf 91}, 216
(2001); {\bf 100}, 197 (2001) 
\bibitem{5} K. Eguchi et al., the KamLAND collaboration, Phys. Rev Lett., 
{\bf{90}} 021802 (2003); S. Abe et al., The KamLAND collaboration, Phys. Rev. 
Lett., {\bf{100}}, 221803 (2008);  A. Suzuki, Prog. Particle and Nucl. Phys.
 {\bf{65}}, 1 (2008); I. Shimuzu, Nucl. Phys. {\bf{B188}}, 84 (2009)
\bibitem {6} A. Habig, arXiv 1004.2647; P. Adamson et al, the MINOS 
collaboration, arXiv 1007.2791
\bibitem{7} A.A. Aguilar-Arevalo et al., The MiniBooNE Collaboration,
Phys. Rev. Lett. {\bf 103}, 061802 (2009)
\bibitem{8} L. B. Auerbach et al., LSND Collaboration, Phys. Rev {\bf{D72}}, 
076004 (2005)
\bibitem{kdcss09} G. Karagiorgi, Z. Djurcic, J.M. Conrad, M.H. Shaevitz, and
M. Sorel, Phys. Rev. {\bf D 80}, 073001 (2009)
\bibitem{ppw05} H.Pas, S. Pakvasa, and T.J. Weller, Phys. Rev. {\bf D 72}, 
095017 (2005)
\bibitem{ms07} M. Mattoni and T. Schwetz, Phys. Rev. {\bf D 76}, 093005 (2007) 
\bibitem{nw08} A.E. Nelson and E. Walsh, Phys. Rev. {\bf D 77}, 033001 (2008)
\bibitem{mini} The MiniBooNE Collaboration, Phys. Rev. Lett. {\bf 105},
181801 (2010)
\bibitem{minos} The Minos Collaboration, Phys. Rev. Lett. {\bf 103}, 261802 
(2009); Phys. Rev. {\bf D 81},052004 (2010)
\bibitem{jhf} The JHF-Kamioka neutrino Project, arXiv:hep-ex/106019
\bibitem{chooz} CHOOZ Collaboration, M. Apollonio $et\; al$ Eur. J. {\bf C 27},
 331 (2003)

\end{thebibliography}
\end{document}